\documentclass[11pt,a4paper]{article}
\usepackage{jheppub}

\usepackage{amsfonts}
\usepackage{amsthm}
\usepackage{slashed}
\usepackage{graphicx}
\usepackage{mathrsfs,bbm}
\usepackage{subfig}
\usepackage{booktabs}
\usepackage[numbers,sort&compress]{natbib}
\usepackage{lscape}
\usepackage{breqn}
\usepackage{amsmath,amssymb,textcomp,enumerate,alltt,xspace,multirow}
\usepackage{multicol}
\usepackage{boxedminipage}
\usepackage{comment}
\usepackage{tabularx}
\usepackage{adjustbox}

\usepackage[colorlinks=true]{hyperref}
\usepackage[dvipsnames]{xcolor}
\hypersetup{
colorlinks=true,
linkcolor=blue,
linktocpage=true,
citecolor=violet
}

\usepackage{mathtools}
\allowdisplaybreaks

\newcommand{\mtil}[1]{#1}

\makeatletter
\g@addto@macro\bfseries{\boldmath}
\makeatother

\theoremstyle{remark}

\DeclareMathOperator{\td}{d}

\newcommand{\iu}{\mathrm{i}}

\newcommand{\VVH}[1][V]{$#1 #1 H$}
\newcommand{\VVHH}[1][V]{$#1 #1 H H$}
\newcommand{\VVV}[1][V]{${#1}{#1}{#1}$}

\newcommand{\imineq}[2]{\vcenter{\hbox{\includegraphics[height=#2ex]{#1}}}}

\newcommand{\tablediag}{12}

\newcommand{\mhh}{m_{HH}}

\title{NLO QCD corrections to the electroweak production of a Higgs boson pair in the quark-antiquark channel}

\date{\today}

\author[a,b,c]{Marco Bonetti,}
\author[b]{Gudrun Heinrich,}
\author[b]{Philipp Rendler,}
\author[d]{William J.\ Torres~Bobadilla}

\affiliation[a]{Institute for Theoretical Physics, University of Tübingen, Auf der Morgenstelle 14, 72076 Tübingen, Germany} 
\affiliation[b]{Institute for Theoretical Physics, Karlsruhe Institute
  of Technology (KIT), Wolfgang-Gaede-Str.~1, 76131 Karlsruhe, Germany}
\affiliation[c]{Institute for Astroparticle Physics, Karlsruhe Institute of Technology (KIT), 76344 Eggenstein-Leopoldshafen, Germany}
\affiliation[d]{Department of Mathematical Sciences, University of Liverpool, Liverpool L69 3BX, U.K.}
 
\emailAdd{marco.bonetti@uni-tuebingen.de}
\emailAdd{gudrun.heinrich@kit.edu}
\emailAdd{philipp.rendler@kit.edu}
\emailAdd{torres@liverpool.ac.uk}

\preprint{{\small KA-TP-04-2026, P3H-26-008}}

\abstract{
Higgs boson pair production in the massless quark-antiquark channel proceeds at leading order (LO) via electroweak boson loops.
We calculate the next-to-leading order QCD corrections to this process. For the corresponding two-loop amplitudes, an analytic representation has been achieved.
Even though the size of this contribution at the level of total cross sections is below 1\% compared to the LO gluon channel, the effect on differential observables can be in the 10\% range and therefore this contribution should be taken into account when comparing to LHC data.

}


\begin{document}

\maketitle

\section{Introduction}
\label{sec:intro}

Higgs boson pair production is an important process at the LHC because it is the only direct way to measure the
trilinear Higgs self‑coupling and thus experimentally probe the shape of the Higgs potential and electroweak symmetry breaking.
Projections for the High-Luminosity LHC reach an accuracy level on the trilinear Higgs coupling of about 30\%~\cite{ATLAS:2025eii} after collecting an integrated luminosity of 3\,ab$^{-1}$. However, this scenario relies on the assumption that the theory uncertainties will be halved by then. Therefore, theoretical predictions for Higgs boson pair production, in particular for the gluon fusion mode that has the largest cross section, need to reach a level of precision where electroweak corrections cannot be neglected.

Higgs boson pair production in gluon fusion is mediated by heavy quark loops at leading order (LO), calculated for the first time in Refs.~\cite{Eboli:1987dy,Glover:1987nx}. As the next-to-leading order (NLO) corrections involve two-loop integrals with several mass scales, the NLO QCD corrections were first calculated in the heavy-top-limit (HTL), rescaled with the top-mass dependent LO amplitude~\cite{Dawson:1998py}.
The NLO QCD corrections including the full top-quark mass dependence became
available later~\cite{Borowka:2016ehy,Borowka:2016ypz,Baglio:2018lrj,Davies:2019dfy,Baglio:2020ini}, increasing the total LO cross section by about 60\%.
The full NLO QCD corrections of Refs.~\cite{Borowka:2016ehy,Borowka:2016ypz} have been combined with a calculation at approximate NNLO, where  the top mass dependence is also included in the real radiation matrix element~\cite{Grazzini:2018bsd}, called NNLO$_{\mathrm{FTapprox}}$. Earlier work has provided NNLO results in the HTL improved by an expansion by powers in $1/m_t^2$~\cite{deFlorian:2016uhr,Grigo:2015dia}.
Matching of the NLO corrections to parton showers has been performed in Refs.~\cite{Heinrich:2017kxx,Jones:2017giv,Heinrich:2019bkc,Bagnaschi:2023rbx,Davies:2025qjr} in the {\sc Powheg-Box-V2} framework, recently also including corrections at approximate NNLO in the {\sc Geneva} framework~\cite{Alioli:2025xcu}.
N$^3$LO corrections~\cite{Chen:2019lzz,Chen:2019fhs} and N$^3$LO+N$^3$LL corrections~\cite{AH:2022elh} in the heavy-top-limit are also known.
The N$^3$LO results have a residual scale uncertainty of about 3\%. This means that other uncertainties, such as
top mass renormalisation scheme uncertainties or electroweak (EW) corrections gain increasing relevance in the uncertainty budget. 
Currently, the top-mass renormalisation scheme uncertainties are the largest uncertainties for this process~\cite{Baglio:2020wgt,Bagnaschi:2023rbx}, they are estimated to be of the order of 20\%.
However, the large logarithms responsible for the differences between $\overline{\mathrm{MS}}$ and on-shell schemes for the top-quark mass at high energies have been identified~\cite{Jaskiewicz:2024xkd}. Furthermore, 
partial three-loop results have been obtained recently~\cite{Davies:2023obx,Davies:2024znp,Davies:2025ghl,Hu:2025hfc}, where in Ref.~\cite{Davies:2025ghl} it is shown that the mass scheme dependence is reduced significantly at this order.

With these prospects the contributions to Higgs boson pair production of EW type are a source of uncertainty which urgently needs to be reduced, in particular as it is known that EW corrections can significantly affect the shapes of observables.
First partial NLO EW corrections to Higgs boson pair production have been calculated in Refs.~\cite{Borowka:2018pxx,Muhlleitner:2022ijf,Davies:2022ram}. 
The possibility to constrain the quartic Higgs boson coupling indirectly through (partial) EW corrections to Higgs boson pair production has been explored in Refs.~\cite{Borowka:2018pxx,Bizon:2024juq,Ding:2026qto}.
The Yukawa and Higgs boson self-coupling corrections have been calculated in Ref. \cite{Heinrich:2024dnz} using a numerical approach, and in Ref.~\cite{Davies:2025wke} based on a high-energy expansion. The EW corrections involving light quarks have been calculated analytically in Ref.~\cite{Bonetti:2025vfd}.
Top-Yukawa- and light-quark-mediated EW corrections have been presented in Ref.~\cite{Bhattacharya:2025egw}.
The NLO EW corrections in the large top-quark mass expansion up to $1/m_t^8$ have been calculated in Ref.~\cite{Davies:2023npk}, the NLO EW corrections with full $m_t$-dependence have been calculated in Ref.~\cite{Bi:2023bnq}, finding that the NLO EW corrections decrease  the total cross section by about $4$\% and distort the Higgs boson pair invariant mass ($\mhh$) distribution close to the Higgs pair production threshold and at high energies. 

In this work, we present the calculation of the mixed QCD-EW contributions to Higgs boson pair production, where at LO the Higgs bosons are produced by light quarks in the initial state through couplings to electroweak bosons, while the QCD corrections involve real and virtual gluons. 
Due to the large gluon luminosity at the LHC, this contribution is not expected to be large. However, we will show that the effects close to the Higgs pair production threshold on the shape of the Higgs boson pair invariant mass distribution are relatively large. As this is a kinematic region that is very sensitive to deviations of the trilinear coupling from its SM value, it is important to include all SM contributions that may affect the low-$\mhh$ region at percent level. The corresponding code will be made publicly available within the \textsc{Powheg-Box-V2}~\cite{Alioli:2010xd} framework.

~

This paper is organised as follows. In Section~\ref{sec:ampli}, we describe the structure of the amplitudes and the setup of our calculation. Section~\ref{sec:MIs} is dedicated to the analytic calculation of the one- and two-loop form factors. In Section~\ref{sec:resu} we present phenomenological results, both at total cross section level and for differential observables, before we conclude in Section~\ref{sec:conclu}.

\section{Scattering amplitudes}
\label{sec:ampli}

\subsection{General structure}

We consider the production of two Higgs bosons from a quark and an antiquark, where the Higgs bosons couple to vector bosons, see Table~\ref{tab:amplqq} for representative diagrams. At LO, this is an electroweak process and we calculate NLO QCD corrections to it.
We use the ``all incoming" convention to label the momenta, therefore we have 
\begin{align}
    \label{eq:qqHH}
    \overline{q}(p_1) \, q(p_2) \to H(-p_3) \, H(-p_4)
\end{align}
for the LO and the NLO virtual contributions, and we use $p_5$ for the additional gluon momentum present in the real radiation diagrams. We use four incoming light flavours, $q \in \{u,d,s,c\}$, with $p_1^2 = p_2^2 = 0$, and consider the Higgs bosons and the gluon to be on-shell ($p_3^2 = p_4^2 = m_H^2$, $p_5^2 = 0$).

As we only consider massless quarks, no Yukawa couplings to the Higgs boson are present. As a consequence, the Higgs bosons can only be generated through vertices with two electroweak (or Goldstone) bosons, resulting in the LO being already a loop-mediated contribution. 
We neglect bottom quarks in the initial state, as they are suppressed by the parton distribution functions (PDFs) relative to the lighter quarks.
Assuming a diagonal CKM matrix, this implies that no top quarks are contained in the two-loop diagrams.
We discuss the impact of bottom quark contributions in Sec.~\ref{sec:resu}.

All diagrams considered here can be classified according to the production mode of the Higgs boson pair, see Table~\ref{tab:amplqq}
and Ref.~\cite{Bonetti:2025vfd}.\footnote{We adopt here a slightly different notation w.r.t.~\cite{Bonetti:2025vfd}, oriented at the coupling structures rather than the propagators. For comparison, $g_{VVHH}$ here corresponds to \VVHH{} of \cite{Bonetti:2025vfd}, $g_{VVH}g_3$ to \VVH{}, and $g_{VVH}^2$ to \VVV{}.
}

We start by considering \emph{bridge diagrams}, where a gauge boson line acts a bridge connecting two 1PI parts of a diagram. These diagrams are zero either by colour (when a gluon acts as a bridge) or by angular momentum conservation (when a photon or a $Z$ boson connects the two parts of the diagram).

The remaining diagrams can be classified into three categories: 
diagrams proportional to $g_{VVHH}$, containing a four-point vertex connecting two EW bosons and two Higgs bosons, 
diagrams proportional to $g_{VVH}\,g_3$, where two EW bosons merge into an $s$-channel Higgs boson subsequently splitting into two, thus containing the trilinear Higgs coupling $g_3$, and 
diagrams proportional to $g_{VVH}^2$, where each Higgs boson is produced from a different $VVH$ three-point vertex. In all these cases, each individual diagram contains either $W$ or $Z$ bosons only (plus the corresponding Goldstone bosons). Each of these three categories is gauge-invariant and closed under renormalisation and IR subtraction.\footnote{We explicitly computed diagrams in both unitary and Feynman gauge, finding identical expressions for each category.}

Due to angular momentum conservation, the diagrams proportional to $g_{VVHH}$ and $g_{VVH}\,g_3$ are zero for LO and virtual NLO contributions, leaving only diagrams of $g_{VVH}^2$-type to contribute at LO and at two-loop order.

\begin{table}
\centering
    \caption{Example diagrams for different classes of contributions to $q\overline{q} \to HH\,(+g)$ at LO, virtual NLO, and real NLO. Each class features diagrams containing either only $W$ (Goldstone) bosons, or only $Z$ (Goldstone) bosons.}
    \label{tab:amplqq}
    \begin{tabular}{ccccc}
        \toprule
        & $g_{VVHH}$ & $g_{VVH}\,g_3$ & $g_{VVH}^2$ & Bridge\\
        \midrule
        LO   &$\imineq{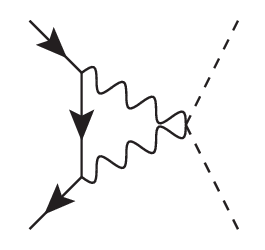}{\tablediag}$  &$\imineq{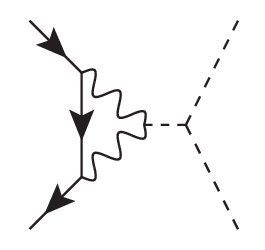}{\tablediag}$  &$\imineq{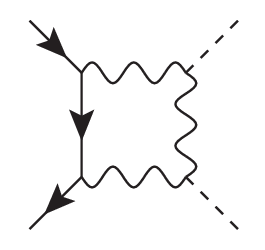}{\tablediag}$ &$\imineq{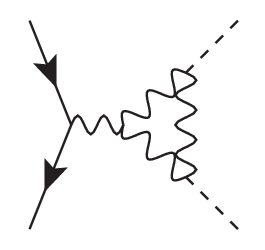}{\tablediag}$  \\
        vNLO &$\imineq{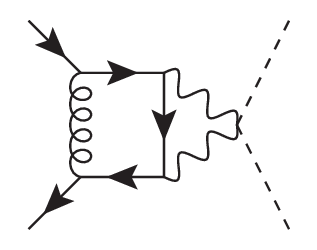}{\tablediag}$ &$\imineq{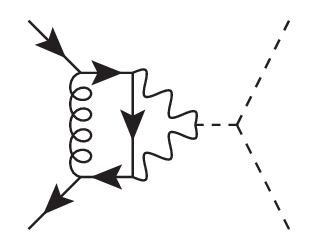}{\tablediag}$ &$\imineq{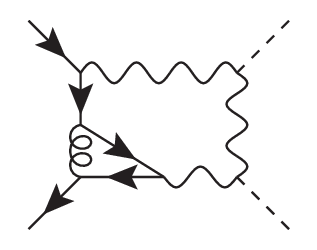}{\tablediag}$ &$\imineq{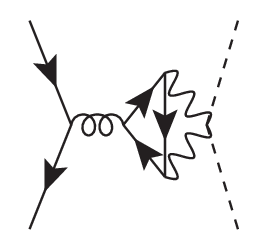}{\tablediag}$\\
        rNLO &$\imineq{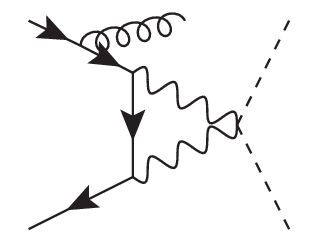}{\tablediag}$  &$\imineq{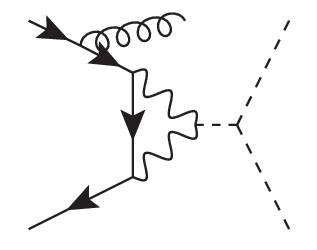}{\tablediag}$  &$\imineq{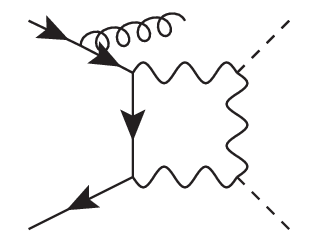}{\tablediag}$ &$\imineq{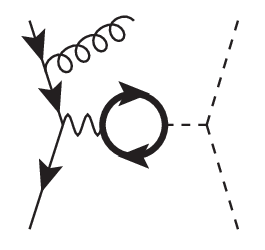}{\tablediag}$ \\
      \bottomrule
    \end{tabular}
\end{table}

$W$ and $Z$ bosons interact with quarks through vertices of the form $\gamma^\mu(g_L \mathbb{P}_L + g_R \mathbb{P}_R)$, potentially introducing traces of gamma matrices containing $\gamma_5$ and resulting in Levi-Civita symbols. We avoid this by considering polarised states for the external quarks, according to the following procedure.

Up to two loops, $W$ and $Z$ bosons are always attached to the same open fermion line, therefore we can anticommute all $\mathbb{P}_{L,R}$ projectors until they touch the rightmost spinor ($u_2$) in the $\overline{u}_1 \gamma\cdots\gamma u_2$ chain.\footnote{Since we are working in dimensional regularisation, $\gamma_5$ is not properly defined, and requires a scheme choice to be consistently handled. We rely on the Kreimer scheme~\cite{Kreimer:1989ke,Korner:1991sx}.} The result is equivalent to selecting a specific polarisation of the external states, starting from a theory where $W$ and $Z$ bosons couple only vectorially to the fermions, later characterising specific polarisation configurations by different couplings. Accordingly, we treat the amplitude as free from $\gamma_5$ terms and reinstate them only at the very end, when considering different helicity configurations.

A potential risk in this procedure is that some of the diagrams might require an even number of anticommutations to move the $\mathbb{P}_{L,R}$ projectors to the rightmost end of the spinor chain, while some other an odd number. As a result, some diagrams with the axial terms stripped off might be equal to \emph{minus} the corresponding diagram with vector couplings only, while some others to \emph{plus} the corresponding diagram, not allowing us to calculate the amplitude as a whole starting from vector-only couplings and then extract the single helicity amplitudes.
However, this is not the case: we are considering massless quarks and QCD corrections, therefore the $q\overline{q}V$ vertex closest to $u_2$ can either be already attached to it or have $n$ $q\overline{q}g$ vertices and massless quark propagators in between, requiring in both cases $2n$ anticommutations to bring the $\mathbb{P}_{L,R}$ next to $u_2$,  therefore producing no relative minus sign for any of the diagrams.
 
As a consequence, the sum of the resulting ``vector-type" diagrams remains consistent with the sum of the original diagrams.

We can now write the $q\overline{q} \to HH$ and $q\overline{q} \to HHg$ amplitudes, respectively, as
\begin{align}
    \begin{aligned}
        M_{i_1 i_2}^{s_1 s_2,q_1 q_2} &= \delta_{i_1 i_2} \,\,\,\,\,\,\,\,\sum_{V=W,Z} \overline{u}^{s_1}_{q_1}(p_1) \tilde{\mathcal{M}}(m_V^2)\,\,\, [\mtil{G}_{V,L}^{q_1 q_2}\,\mathbb{P}_L + \mtil{G}_{V,R}^{q_1 q_2}\,\mathbb{P}_R] u^{s_2}_{q_2}(p_2)
        \,, \\
        M_{c,i_1 i_2}^{s_1 s_2\lambda,q_1 q_2} &= \left(T_c\right)_{i_1 i_2} \sum_{V=W,Z} \overline{u}^{s_1}_{q_1}(p_1) \tilde{\mathcal{M}}^\mu(m_V^2) [\mtil{G}_{V,L}^{q_1 q_2}\,\mathbb{P}_L + \mtil{G}_{V,R}^{q_1 q_2}\,\mathbb{P}_R] u^{s_2}_{q_2}(p_2) \epsilon^{*\lambda}_\mu(p_5)
        \,,
    \end{aligned}
\end{align}
with abuse of notation in the momentum conservation, 
$p_1+\hdots+p_4(+p_5)=0$; $i_1$ and $i_2$ are the antiquark and quark colour indices, respectively, $s_1$ and $s_2$ the spin indices, $q_1$ and $q_2$ the flavour indices, $c$ the colour index of the extra gluon, $\lambda$ its polarisation, and $\tilde{\mathcal{M}}$ and $\tilde{\mathcal{M}}^\mu$ the amplitudes where each $q\overline{q}V$ vertex has been replaced with a pure vector interaction, which can be further decomposed, following to Table~\ref{tab:amplqq}, into
\begin{align}
        \tilde{\mathcal{M}}
        =
        \tilde{\mathcal{M}}_{g_{WWH}^2} +
        \tilde{\mathcal{M}}_{g_{ZZH}^2}
\end{align}
and
\begin{align}
    \begin{aligned}
        \tilde{\mathcal{M}}^\mu
        =
        \tilde{\mathcal{M}}^\mu_{g_{WWHH}} +
        \tilde{\mathcal{M}}^\mu_{g_{WWH} g_3} +
        \tilde{\mathcal{M}}^\mu_{g_{WWH}^2} +
        \tilde{\mathcal{M}}^\mu_{g_{ZZHH}} +
        \tilde{\mathcal{M}}^\mu_{g_{ZZH} g_3} +
        \tilde{\mathcal{M}}^\mu_{g_{ZZH}^2}
        \,.
    \end{aligned}
\end{align}

The $G_{V,L/R}^{q_1q_2}$ couplings encapsulate the SM EW coupling structure of the different parts of the amplitude and read~(cf.\ \cite{Romao:2012pq,Denner:2019vbn,Bonetti:2022lrk})
\begin{align}
    &\begin{aligned}
        \mtil{G}_{W,L}^{q_1 q_2} &= 
        \begin{dcases}
            \sum_{k \in \{d,s\}} \mathcal{V}^*_{q_1 k}\mathcal{V}_{k q_2} & \text{if } q_1 \in \{u,c\}  \\
            \sum_{k \in \{u,c\}}   \mathcal{V}_{k q_1}\mathcal{V}^*_{q_2 k} & \text{if } q_1 \in \{d,s\}
        \end{dcases}
    \end{aligned}\\
    &\begin{aligned}
        \mtil{G}_{W,R}^{q_1 q_2} &= 0
    \end{aligned}\\
    &\begin{aligned}
        &\mtil{G}_{Z,L}^{q_1 q_2} = \frac{2\delta_{q_1 q_2}}{c_W^4} \left(g_{L}^{q_1 q_2}\right)^2
        =
        \begin{dcases}
            \frac{1}{c_W^4} \left( \frac{1}{2} - \frac{4}{3} s_W^2 + \frac{8}{9} s_W^4 \right) & \text{if } q_1=q_2 \in \{u,c\}  \\
            \frac{1}{c_W^4} \left(\frac{1}{2} - \frac{2}{3}s_W^2 + \frac{2}{9}s_W^4\right)& \text{if } q_1=q_2 \in \{d,s\}
        \end{dcases}    
    \end{aligned}\\
    &\begin{aligned}
        &\mtil{G}_{Z,R}^{q_1 q_2} = \frac{2\delta_{q_1 q_2}}{c_W^4} \left(g_{R}^{q_1 q_2}\right)^2 
        =
        \begin{dcases}
            \frac{1}{c_W^4} \left( \frac{8}{9} s_W^4 \right) & \text{if } q_1=q_2 \in \{u,c\}  \\
            \frac{1}{c_W^4} \left( \frac{2}{9} s_W^4 \right) & \text{if } q_1=q_2 \in \{d,s\}
        \end{dcases}
    \end{aligned}
\end{align}
($\mathcal{V}$ is the Cabibbo--Kobayashi--Maskawa mixing matrix, $c_W=\cos\theta_W$, $s_W=\sin\theta_W$, and $\theta_W$ is the Weinberg angle).

In terms of helicity amplitudes, we have only two non-zero configurations for the $q\overline{q} \to HH$ case:
\begin{align}
    &\begin{aligned}
        \label{eq:haLR}
        M_{i_1 i_2}^{-+,q_1 q_2} &= 
        \delta_{i_1 i_2} \sum_{V=W,Z} \mtil{G}_{V,R}^{q_1 q_2}\,  [1| \,\tilde{\mathcal{M}}(m_V^2) |2\rangle
        \,,
    \end{aligned}
    \\
    &\begin{aligned}
        \label{eq:haRL}
        M_{i_1 i_2}^{+-,q_1 q_2} &= 
        \delta_{i_1 i_2} \sum_{V=W,Z} \mtil{G}_{V,L}^{q_1 q_2} \langle1| \,\tilde{\mathcal{M}}(m_V^2) |2]
        \,.
    \end{aligned}
\end{align}
Notice that the two configurations, up to $\mtil{G}_{V,L/R}^{q_1 q_2}$ coefficients, are related by complex conjugation of the spinor structures, resulting in only one independent non-zero helicity expression. This is consistent with working with a purely vectorial amplitude, reconstructing the axial contributions at a later stage.

For $q\overline{q} \to HHg$ we obtain four non-zero configurations:
\begin{align}
    &\begin{aligned}
        \label{eq:haLR+}
        M_{i_1 i_2}^{-++,q_1 q_2} &= 
        \left(T_c\right)_{i_1 i_2} \sum_{V=W,Z}\delta_{i_1 i_2}\, \mtil{G}_{V,R}^{q_1 q_2}\,  [1| \,\tilde{\mathcal{M}}^\mu(m_V^2) |2\rangle \frac{[r|\gamma_\mu|5\rangle}{\sqrt{2}[5r]}
        \,,
    \end{aligned}
    \\
    &\begin{aligned}
        \label{eq:haRL+}
        M_{i_1 i_2}^{+-+,q_1 q_2} &= 
        \left(T_c\right)_{i_1 i_2} \sum_{V=W,Z}\delta_{i_1 i_2}\, \mtil{G}_{V,L}^{q_1 q_2} \langle1| \,\tilde{\mathcal{M}}^\mu(m_V^2) |2] \frac{[r|\gamma_\mu|5\rangle}{\sqrt{2}[5r]}
        \,;
    \end{aligned}
    \\
    &\begin{aligned}
        \label{eq:haLR-}
        M_{c,i_1 i_2}^{-+-,q_1 q_2} &= 
        \left(T_c\right)_{i_1 i_2} \sum_{V=W,Z}\delta_{i_1 i_2}\, \mtil{G}_{V,R}^{q_1 q_2}\,  [1| \,\tilde{\mathcal{M}}^\mu(m_V^2) |2\rangle \frac{\langle r|\gamma_\mu|5]}{\sqrt{2}\langle r5\rangle}
        \,,
    \end{aligned}
    \\
    &\begin{aligned}
        \label{eq:haRL-}
        M_{c,i_1 i_2}^{+--,q_1 q_2} &= 
        \left(T_c\right)_{i_1 i_2} \sum_{V=W,Z}\delta_{i_1 i_2}\, \mtil{G}_{V,L}^{q_1 q_2} \langle1| \,\tilde{\mathcal{M}}^\mu(m_V^2) |2] \frac{\langle r|\gamma_\mu|5]}{\sqrt{2}\langle r5\rangle}
        \,,
    \end{aligned}
\end{align}
where $r$ is the reference vector associated to the gluon polarization. Here as well, up to the electroweak prefactors $\mtil{G}_{V,L/R}^{q_1 q_2}$, the $+-+$ and $-+-$ as well as the $-++$ and $+--$ helicity configurations are related by complex conjugation of the spinor structures.

\subsection{LO \& virtual NLO amplitudes}
\label{sec:vNLO}

Once stripped from its axial contributions, the $q\overline{q} \to HH$ amplitude can be written in terms of a single form factor as
\begin{align}
    \overline{u}^{s_1}(p_1) \,\tilde{\mathcal{M}}_{g_{VVH}^2}(m_V^2) u^{s_2}(p_2) &= \overline{u}^{s_1}(p_1) \slashed{p}_3 u^{s_2}(p_2) \, \mathcal{F}_{g_{VVH}^2}(m_V^2)
    \,,
\end{align}
with $V=W,Z$.\footnote{This is again consistent with the existence of a single independent non-zero helicity configuration, \\see \cite{Peraro:2019cjj,Peraro:2020sfm}.} The bare form factor $\mathcal{F}_{g_{VVH}^2}(m_V^2)$ admits the decomposition
\begin{align}
    \label{eq:dec}
    \mathcal{F}_{g_{VVH}^2}(m_V^2) = g_{VVH}^2 \omega_{q}(m_V^2) \sum_{L=1}^\infty a^{L-1}(m_V^2) \mathcal{A}^{(L)}(m_V^2)
    \,,
\end{align}
with $g_{VVH} = 1$ in the SM and
\begin{align}
    \label{eq:omegaa}
    \omega_q(m_V^2) &= \iu (4\pi)^\epsilon \mathrm{e}^{-\gamma_E \epsilon} \left(\frac{m_V^2}{\mu_{\textup{EW}}^2}\right)^{-\epsilon} \frac{\alpha^2}{m_V^2 \sin^4 \theta_W}
    \,,\\
    a(m_V^2) &= (4\pi)^\epsilon \mathrm{e}^{-\gamma_E \epsilon} \left(\frac{m_V^2}{\mu^2}\right)^{-\epsilon} \left(\frac{\alpha_S}{2\pi}\right)
    \,,
\end{align}
where $\mu_{\textup{EW}}^2$ and $\mu^2$ are the regularisation parameters coming from loop integrals---the first one from the LO EW couplings, the second one from higher-order QCD corrections. We define the Mandelstam invariants as
\begin{align}
    \begin{aligned}
        s &= (p_1+p_2)^2
        \,,\\
        t &= (p_1+p_3)^2
        \,,\\
        u &= (p_2+p_3)^2
        \,,
    \end{aligned}
\end{align}
obeying $s + t + u = 2 m_H^2$, with momenta defined as in Eq.~\eqref{eq:qqHH}.

The form factor $\mathcal{F}_{g_{VVH}^2}(m_V^2)$ can be extracted by applying the projector
\begin{align}
    \mathbb{P}^{s_2 s_1} = - \frac{\overline{u}^{s_2}(p_2) \slashed{p}_3 u^{s_1}(p_1)}
    {2 (m_H^4 - t u)}
\end{align}
to the amplitude as
\begin{align}
    \mathcal{F}_{g_{VVH}^2}(m_V^2) = \overline{u}^{s_1}(p_1) \,\tilde{\mathcal{M}}_{g_{VVH}^2}(m_V^2) u^{s_2}(p_2) \mathbb{P}^{s_2 s_1}
    \,.
\end{align}

We extract the virtual NLO finite remainder following the procedure described in~\cite{Catani:1998bh}. We first employ the $\overline{\text{MS}}$ renormalisation scheme for the strong coupling
\begin{align}
    \label{eq:ren}
    \alpha_S = \frac{1}{S_\epsilon}\left(\frac{\bar{\mu}^2}{\mu^2}\right)^\epsilon \overline{\alpha_S}(\bar{\mu}^2) \left[1+\mathcal{O}(\overline{\alpha_S})\right]
    ,
\end{align}
where $\alpha_S$ is the bare coupling, $\overline{\alpha_S}(\bar{\mu}^2)$ is the renormalised one, $\bar{\mu}^2$ is the renormalisation scale, and $S_\epsilon = \mathrm{e}^{-\gamma_E \epsilon}(4\pi)^\epsilon$. We then apply the operator
\begin{align}
    \begin{aligned}
        \label{eq:Cat1}
        \mathbf{I}_1 
        = - C_F \left(\frac{2}{\epsilon^2} + \frac{3}{\epsilon}\right) \frac{\mathrm{e}^{\gamma_E \epsilon}}{2\Gamma(1-\epsilon)}\left(-\frac{\bar{\mu}^2}{s}\right)^\epsilon
        ,
    \end{aligned}
\end{align}
with $C_F=(N_C^2-1)/(2N_C)$, to extract the finite remainder as
\begin{align}
    \mathcal{A}^{(2)}_{\text{fin}} &= \mathcal{A}^{(2)} - \left(\frac{m_V^2}{\bar{\mu}^2}\right)^\epsilon \mathcal{A}^{(1)} \mathbf{I}_1
        \,.
\end{align}

\subsection{Real NLO amplitudes}

Real corrections are provided by the process $q\overline{q} \to H H g$, together with its permutations $q g \to H H q$ and $\overline{q} g \to H H \overline{q}$, where the configuration $g_{VVHH} = g_{VVH} = g_3 = 1$ corresponds to the SM. Contrary to what was observed for LO and virtual corrections, 
contributions proportional to $g_{VVHH}$ and $g_{VVH}\,g_3$ are non-zero here, thanks to the angular momentum carried by the extra gluon.

While the $g_{VVH}^2$ contributions will present standard NLO soft and collinear divergences due to the extra gluon, the novel $g_{VVHH}$ and $g_{VVH} g_3$ contributions are finite in the same phase-space regions, due to the corresponding LO contributions being zero. This can be seen already at the diagram level in the soft limit by employing the soft Feynman rules presented in \cite{Catani:2000pi} and in the collinear limit performing a similar analysis.

We automatically generate the real emission contributions using the computer code \textsc{GoSam}~\cite{Cullen:2011ac,Cullen:2014yla,Braun:2025afl}.

\section{Master Integrals}
\label{sec:MIs}

\subsection{Mathematical framework}

\begin{figure}
\centering
    \subfloat[]{{\includegraphics[width=0.33\textwidth]{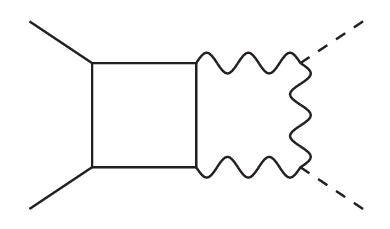}}\label{fig:PL}}
    \quad
    \subfloat[]{{\includegraphics[width=0.33\textwidth]{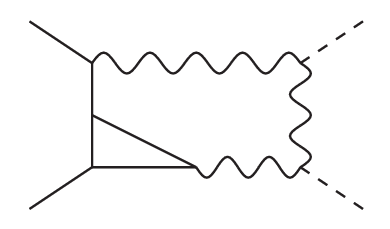}}\label{fig:PL2}}
	\caption{Two-loop top sectors. Massless particles are indicated by straight lines, particles with mass $m_V$ by wavy lines, and particles with mass $m_H$ by dashed lines.}
    \label{fig:MI}
\end{figure}

For the analytic evaluation of the form factors at one and two loops, we adopt the standard methodology used for multi-loop scattering amplitudes. We begin by reducing the relevant Feynman integrals to a minimal set of master integrals using integration-by-parts (IBP) identities~\cite{Chetyrkin:1981qh,Laporta:2000dsw}, implemented through {\sc Reduze}~\cite{fermat,vonManteuffel:2012np} and {\sc Kira}~\cite{Maierhoefer:2017hyi,Klappert:2019emp,Klappert:2020nbg,Klappert:2020aqs}.
We then construct a basis of master integrals that satisfy a canonical system of differential equations~\cite{Henn:2013pwa} using {\sc LiteRed}~\cite{Lee:2012cn} and {\sc FiniteFlow}~\cite{Peraro:2019svx}. At one loop, we identify 11 canonical master integrals, determined by analysing their leading and Landau singularities, along the lines of~\cite{Flieger:2022xyq}. 
At two loops, only planar integrals are required for the computation of the form factor $\mathcal{A}^{(2)}$, specifically those illustrated in Fig.~\ref{fig:MI}. 
All master integrals associated with the double-box topology shown in Fig.~\ref{fig:PL} were recently computed in~\cite{Bonetti:2025vfd}. 
To complete the canonical basis, only six additional integrals---derived from the topology in Fig.~\ref{fig:PL2}---are needed. 
We identify the missing integrals with the help of {\sc DlogBasis}~\cite{Henn:2020lye}.
Let us remark that the integration kernels in the canonical differential equations (or letters of the kinematic alphabet)
are contained in the alphabet reported in~\cite{Bonetti:2025vfd}. 
Hence, no new letters arise from the topology in Fig.~\ref{fig:PL2}.

Once the complete set of master integrals is established, we rotate the integral basis to a basis 
of graded transcendental functions, as described in~\cite{Chicherin:2021dyp,Gehrmann:2024tds}. 
This functional organisation not only simplifies the analytic structure of the result but also ensures 
efficient cancellation of IR poles when computing the finite remainder at two loops. 
In particular, we incorporate the Catani subtraction operator~\eqref{eq:Cat1} 
to handle the IR structure of the one-loop amplitude, and we ensure that the transcendental functions 
are organised in a way that allows for 
a matching of transcendental functions 
of the leading and subleading $\epsilon$ poles at two loops. 
This is achieved by identifying relations between products of logarithms and the independent transcendental functions. 
For the remaining integrals, boundary constants are calculated in the limit $s,t,u\ll m_{V}^{2}$, often referred to
as the large-mass expansion, in the same way as carried out in~\cite{Bonetti:2025vfd}.

For the sake of reproducibility and to facilitate further analysis, 
we provide supplemental material at~\cite{bonetti_2026_18285680} with 
the full expressions for the one- and two-loop bare amplitudes $\mathcal{A}^{(1)}$ and $\mathcal{A}^{(2)}$, 
expanded up to $\mathcal{O}(\epsilon^2)$ and $\mathcal{O}(\epsilon^0)$, respectively, as well as the finite remainder $\mathcal{A}^{(2)}_{\text{fin}}$. 
These expressions are given in terms of our chosen basis of independent transcendental functions
together with their explicit representation as Chen iterated integrals.

\subsection{Amplitude evaluation}

We follow the same strategy as outlined in~\cite{Bonetti:2025vfd} to provide numerical values, starting either from our basis of independent transcendental functions or from the single Chen iterated integrals. 
Namely, we evolve the boundary point to the desired target point by solving the differential equations 
numerically with the help of {\sc DiffExp}~\cite{Moriello:2019yhu,Hidding:2020ytt}. Our expressions can also be interfaced with 
other codes which support complex kinematics~\cite{Armadillo:2022ugh,Prisco:2025wqs,PetitRosas:2025xhm}.

We begin by studying the analytic structure of the one-loop form factor, $\mathcal{A}^{(1)}$,
expressed in terms of independent transcendental functions, say $w^{(k)}$, with $k$ the transcendental weight. 
This allows us to make manifest the functional relations and simplifications that emerge in the finite remainder, 
and provides insight into the interplay between the algebraic and transcendental structures of the amplitude. 
The finite contribution to the one-loop form factor $\mathcal{A}^{(1)}$ simply becomes:
\begin{align}
\label{eq:A1}
\mathcal{A}_{\text{fin}}^{\left(1\right)}&=\frac{(t-u)(t+u-4)}{4r_{1}\left(m_{H}^{4}-tu\right)}w_{7_{2}}^{\text{(2)}}+\frac{(u-t)}{4\left(m_{H}^{4}-tu\right)}w_{{13}_{2}}^{\text{(2)}}
+\frac{(t-2)}{2\left(m_{H}^{4}-tu\right)}w_{2_{2}}^{\text{(2)}}-\frac{(u-2)}{2\left(m_{H}^{4}-tu\right)}w_{{20}_{2}}^{\text{(2)}}
\notag
\\
&{}+\frac{1}{2r_{3}}\left(1-t+\frac{(t-2)((s-1)t+u)}{2\left(m_{H}^{4}-tu\right)}\right)w_{{10}_{2}}^{\text{(2)}}
\notag\\
&-\frac{1}{2r_{8}}\left(1-u+\frac{(u-2)((s-1)u+t)}{2\left(m_{H}^{4}-tu\right)}\right)w_{{23}_{2}}^{\text{(2)}}\,.
\end{align}
where we set $m_V=1$, and the transcendental functions $\omega^{(2)}$ obey the system of differential equations, 
following the strategy of~\cite{Caron-Huot:2014lda}, 
\begin{align}
    \td\vec{\omega}_{1;0} = \td\Omega_{1;0}\,\vec{\omega}_{1;0}\,,
\end{align}
with,
\begin{align}
\vec{w}_{3:0}=
\big\{ w_{1_{0}}^{(0)},w_{1_{1}}^{(1)},w_{2_{1}}^{(1)},w_{5_{1}}^{(1)},w_{7_{1}}^{(1)},w_{2_{2}}^{(2)},
w_{7_{2}}^{(2)},w_{10_{2}}^{(2)},w_{13_{2}}^{(2)},w_{20_{2}}^{\text{(2)}},w_{23_{2}}^{\text{(2)}}\big\}\,,
\end{align}
and the only non-vanishing boundary value at $s=0$, 
\begin{align}
w_{1_{0}}^{\text{(0)}}\big|_{s=0} = -1\,,
\end{align}
In Appendix~\ref{app:A}, we provide the explicit form of the differential equations satisfied by the functions appearing in Eq.~\eqref{eq:A1}, along with precise definitions of the square roots $r_1, r_3, r_8$ 
that enter the alphabet of the system. 
For a more detailed discussion of the underlying method, we refer the reader to~\cite{Bonetti:2025vfd}.

Since the one- and two-loop canonical integrals are both expressed in terms of the same set of transcendental functions, we construct a unified system of 152 differential equations for all functions appearing in 
$\mathcal{A}^{(1)}$ and $\mathcal{A}^{(2)}$. 
Although these form factors depend on a smaller subset---92 independent transcendental functions---additional equations are required to close the system of differential equations.
These differential equations are organised according to their transcendental weight, 
which reflects the number of iterated integrations. 

\begin{table}[t]
    \centering
    \caption{Real and imaginary parts of the numerical evaluation of the LO ($\mathcal{A}^{(1)}$), bare virtual NLO ($\mathcal{A}^{(2)}$), and NLO finite remainder ($\mathcal{A}^{(2)}_{\text{fin}}$) amplitudes at the phase-space point of Eq.~\eqref{eq:psp_test}.}
    \label{tab:FF_num_eval}
    \begin{tabular}{lcrcr}
    \toprule
    $~\mathcal{A}^{(1)},\epsilon^{0}$&~     &$0.11720874553989$&$-$
                        &$0.11116963414362\,\iu~$\\
    $~\mathcal{A}^{(1)},\epsilon^{1}$&~     &$0.06744433484893$&$+$
                        &$0.15073003495756\,\iu~$\\
    $~\mathcal{A}^{(1)},\epsilon^{2}$&~     &$-0.100790286461074$&$+$
                        &$0.013669915468337\,\iu~$\\
    \midrule
    $~\mathcal{A}^{(2)},\epsilon^{-2}$&~   &$-0.156278327386515$&$+$
                                    &$0.148226178858166\,\iu~$    \\
    $~\mathcal{A}^{(2)},\epsilon^{-1}$&~   &$-0.290675394584724$&$-$
                                    &$0.943203210511806\,\iu~$    \\
    $~\mathcal{A}^{(2)},\epsilon^{0}$&~    &$1.666836281405315$&$+$
                                    &$0.560490323962583\,\iu~$    \\
    \midrule
    $~\mathcal{A}^{(2)}_{\text{fin}}$&~    &$-0.891755853165636$&$+$
                                    &$0.495448608356593\,\iu~$    \\
    \bottomrule
    \end{tabular}
\end{table}

We numerically evaluate the one- and two-loop form factors
at the following phase-space point in the physical region
\begin{align}
    \label{eq:psp_test}
\left\{ s_{0}\,,t_{0}\,,u_{0}\,,m_{V;0}^{2}\right\} 
=
\left\{ \frac{3125}{128}\,,-\frac{1875}{128}\,,-\frac{625}{128}\,,1\right\}\,.
\end{align} 
In Table~\ref{tab:FF_num_eval}, we present the numerical values of the $\epsilon$ expansions for the form factors 
$\mathcal{A}^{(1)}$ and $\mathcal{A}^{(2)}$, as well as the 
two-loop finite remainder $\mathcal{A}^{(2)}_{\text{fin}}$. 

To validate our implementation, we compare the numerical evaluation of the canonical integrals 
appearing in the form factors with results obtained using {\sc AMFlow}~\cite{Liu:2017jxz,Liu:2022chg}, 
finding agreement up to about 20 digits.

\section{Phenomenological results}
\label{sec:resu}

We combine the LO and NLO (real and virtual) amplitudes using the \textsc{Powheg-Box-V2} framework~\cite{Nason:2004rx,Frixione:2007vw,Alioli:2010xd} combined with \textsc{GoSam-3.0}~\cite{Braun:2025afl}.
We calculate the total cross sections at proton-proton centre-of-mass energies of $\sqrt{S} = 13\,\textup{TeV}$, $\sqrt{S} = 13.6\,\textup{TeV}$ and $\sqrt{S} = 14\,\textup{TeV}$, using the \texttt{PDF4LHC21_40} parton distribution function (PDF) set \cite{PDF4LHCWorkingGroup:2022cjn}. For our differential results we use 
 $\sqrt{S} = 13.6\,\textup{TeV}$. As electroweak input parameters we choose the gauge boson masses $m_W=80.36\,\textup{GeV}$, $m_Z=91.1876\,\textup{GeV}$, together with the Fermi constant ${G_F = 0.116639\times10^{-4}\,\textup{GeV}^{-2}}$. The mass of the Higgs boson is fixed to $125.0\,\textup{GeV}$, and the widths of the gauge bosons are neglected. The renormalisation and factorisation scales are dynamically chosen to $\mu_F = \bar{\mu} = m_{HH}/2$. Scale uncertainties for the quark-antiquark channel are determined using a 7-point variation, varying $\bar{\mu}$ and $\mu_F$ independently by factors of $1/2$ and $2$, while omitting the combinations $(\bar{\mu}, \mu_F) = (\mhh/4, \mhh)$ and $(\mhh, \mhh/4)$.

In Table~\ref{tab:crosssection}, we report the total cross sections for the quark-antiquark channel at LO and NLO  as well as the LO contribution from the gluon channel. 
The large K-factor between LO and NLO can be understood from the fact that the gluon radiation opens up new partonic channels, such as $qg$, which are less suppressed by the PDFs than the $q\bar{q}$ channel.

\begin{table}
    \centering
    \caption{Total LO and NLO cross sections at proton-proton centre-of-mass energies of $13.0\,\textup{TeV}$, $13.6\,\textup{TeV}$ and $14.0\,\textup{TeV}$. 
     }
    \label{tab:crosssection}
    \begin{tabular}{cccccc}
    \toprule
    $\sqrt{S}\,[\textup{TeV}]$ & $q\overline{q}_{\text{LO}}\,[\textup{fb}]$ & $q\overline{q}_{\text{NLO}}\,[\textup{fb}]$ & $gg_{\text{LO}}\,[\textup{fb}]$ & $q\overline{q}_{\text{NLO}}/q\overline{q}_{\text{LO}}$ & $q\overline{q}_{\text{NLO}}/gg_{\text{LO}}$\\ 
    \midrule 
    $13.0$ & $0.039$ & $0.061$ & $16.45$ & $+59\%$ & $+0.37\%$\\
    $13.6$ & $0.041$ & $0.066$ & $18.26$ & $+60\%$ & $+0.36\%$\\
    $14.0$ & $0.043$ & $0.069$ & $19.52$ & $+60\%$ & $+0.35\%$\\
    \bottomrule
    \end{tabular}
\end{table}

\subsection{Differential distributions}
\begin{figure} 
\centering 
\includegraphics[scale=0.59]{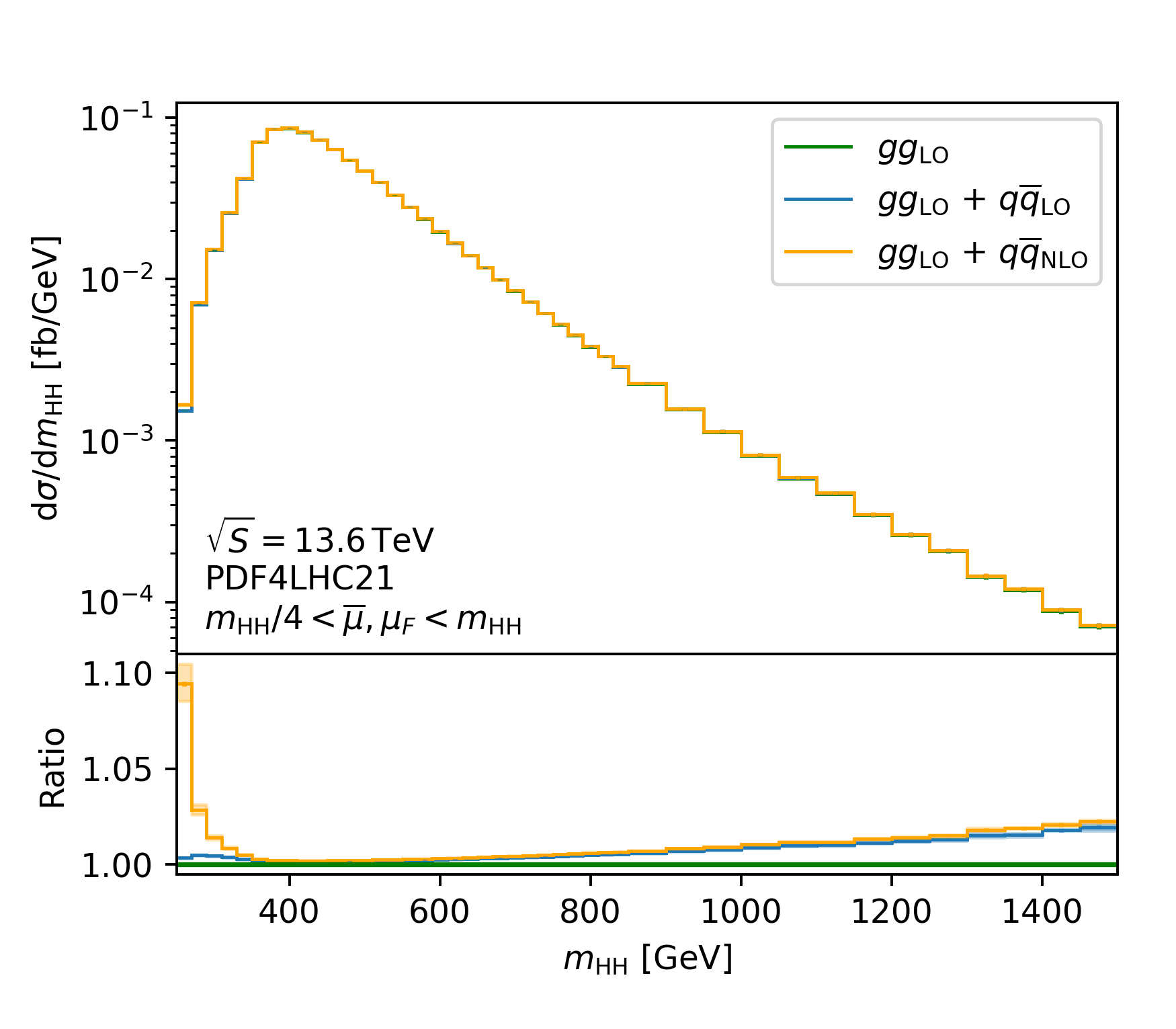}
\includegraphics[scale=0.59]{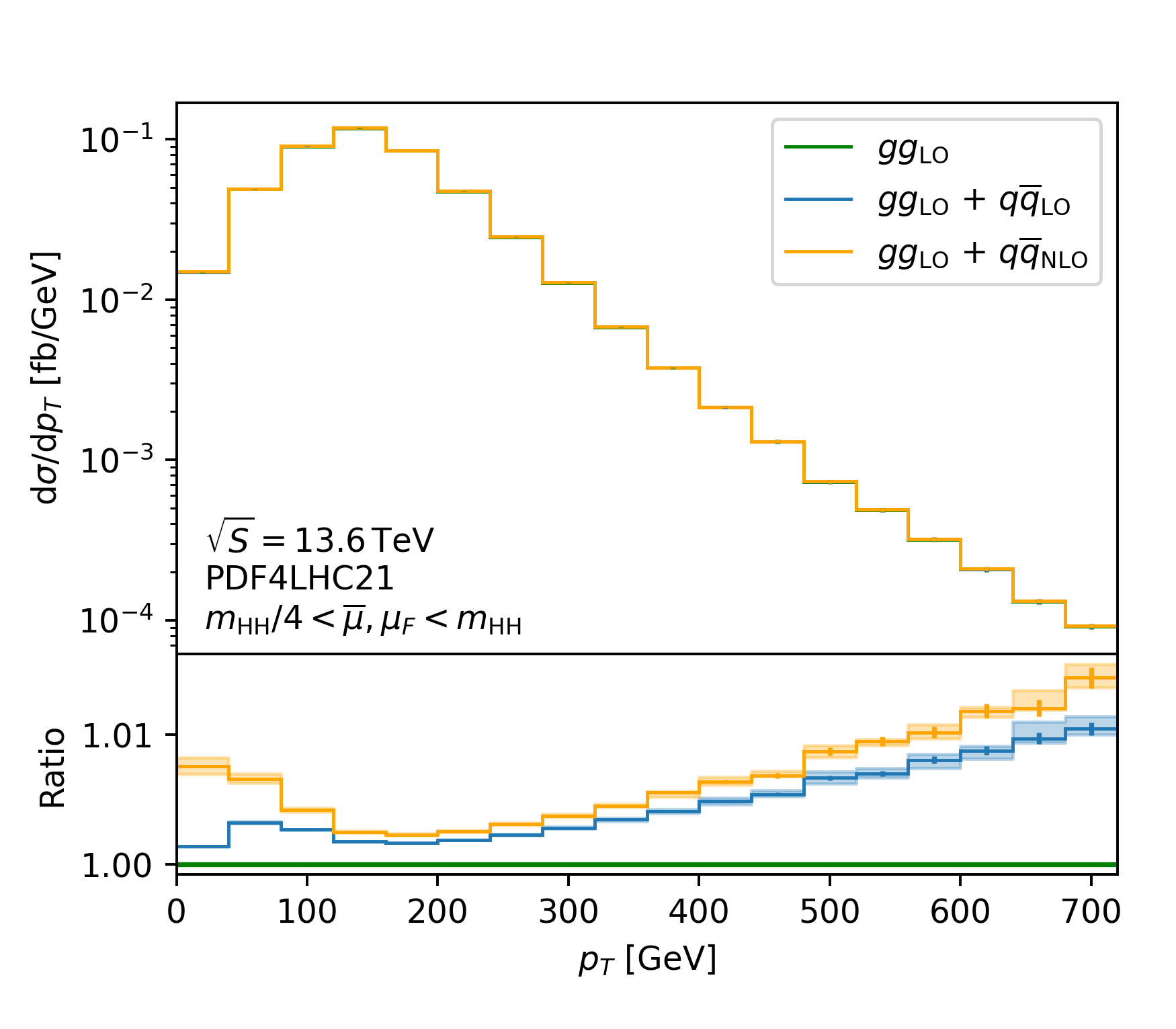}
\includegraphics[scale=0.59]{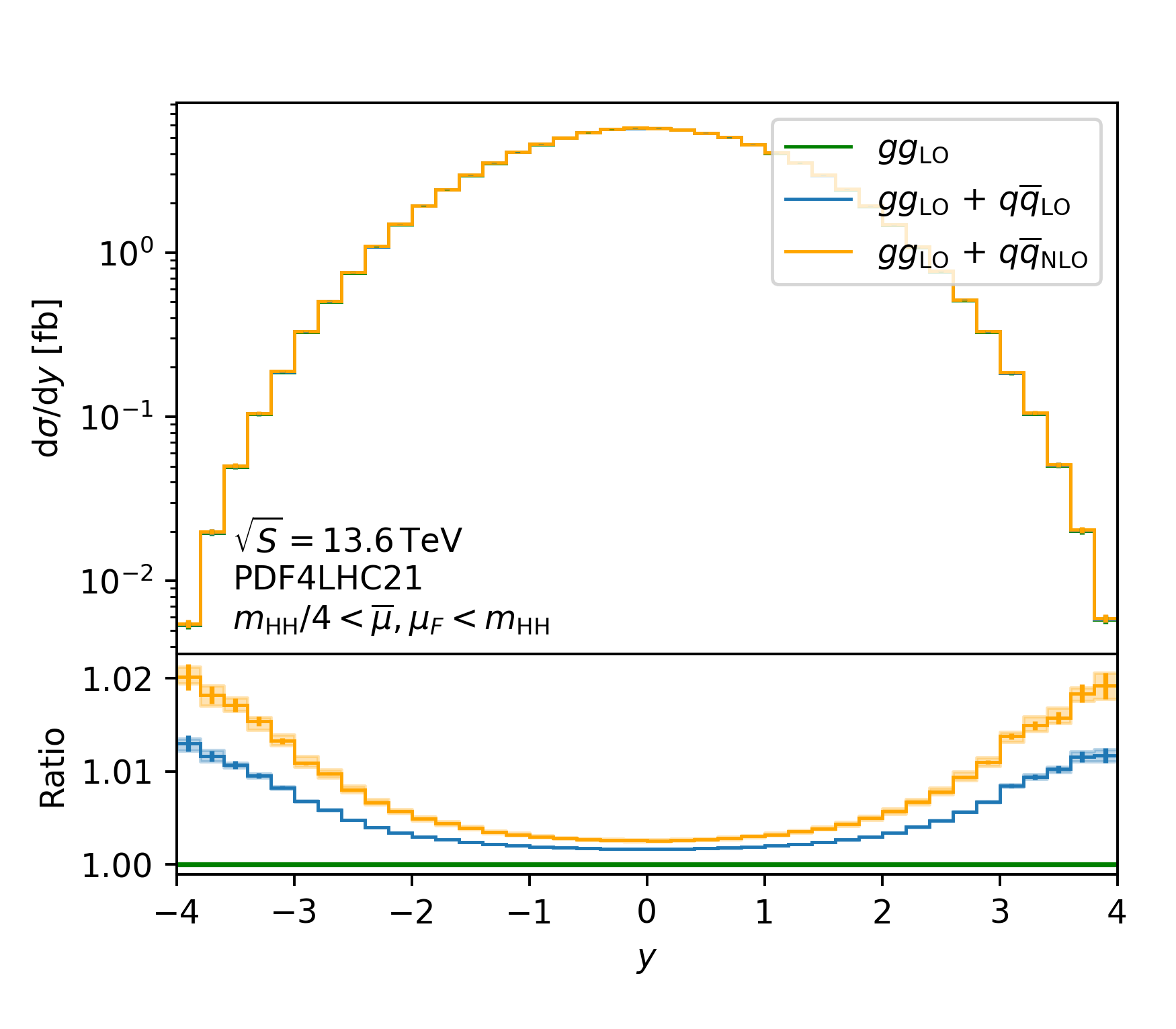}
\caption{Invariant mass distribution of the Higgs pair (top left), transverse momentum distribution (top right) and rapidity distribution (bottom) of a single Higgs boson at a proton-proton centre-of-mass energy of $\sqrt{S} = 13.6\,\textup{TeV}$. The error bands represent the scale uncertainties of the quark-antiquark channel and the error bars indicate the statistical uncertainties from the Monte Carlo integration.
}
\label{fig:invariantMass}
\end{figure}

We provide differential distributions for the invariant mass of the Higgs boson pair, $m_{HH}$, the transverse momentum of one Higgs boson, $p_T$, and the rapidity $y$ of one Higgs boson, shown in Figure~\ref{fig:invariantMass} (top left, top right, and bottom, respectively). For comparison, the LO (blue) and NLO (orange) contributions from the quark-antiquark channel are shown on top of the LO gluon-fusion contribution (green). The upper subplots show the absolute values, while the lower subplots show the relative effect with respect to the gluon channel. 

For the invariant mass distribution (top left), we observe an enhancement of nearly $+10\%$ in the leftmost bin, which is largely driven by real emission contributions and is almost absent at LO. In this kinematic region, two enhancement mechanisms combine:
\begin{enumerate}
    \item The real emission amplitude is enhanced for $m_{HH}$ values close to the production threshold, corresponding to configurations where a hard jet recoils against the Higgs boson pair. 
    \item The real emission contribution includes initial-state gluon splittings into quark-antiquark pairs, which enhance the differential cross section due to the dominance of gluon PDFs at the LHC.
\end{enumerate}
In addition, the quark-antiquark channel produces an enhancement in the high-$m_{HH}$ tail of the distribution, where logarithms of Sudakov-type, $\log{(\hat{s}/m_V^2)}$, grow large.

For the real radiation diagrams, we observe destructive interference between triangle-type and box-type diagrams (cf.~the last line in Table~\ref{tab:amplqq}), similar to the gluon fusion case, while for the virtual corrections this behaviour is absent since the triangle-type diagrams vanish. Therefore we anticipate a large sensitivity to modifications of the Higgs couplings (changing the interference pattern) from the real radiation contributions. 

The transverse momentum distribution (top right) of one of the Higgs bosons exhibits a pattern similar to that observed in the invariant-mass distribution: An enhancement at low $p_T$, primarily driven by real emission contributions, as well as an increase in the high-$p_T$ tail, still modest at $p_T$ values of about 700\,GeV.

The rapidity distribution (bottom) is symmetric around $y = 0$ and nearly vanishes in the central bin, due to the $t \leftrightarrow u$ anti-symmetry of the LO and NLO virtual amplitudes. 

The quark-antiquark channel, together with the NLO electroweak corrections to the gluon channel~\cite{inpreparation}, will be included in the public library \texttt{ggHH}~\cite{Heinrich:2017kxx,Heinrich:2020ckp} within the \textsc{Powheg-Box-V2} framework~\cite{Nason:2004rx,Frixione:2007vw,Alioli:2010xd}, making these results publicly available.

\subsection{Bottom quark effects}
We investigate the effect of initial-state massless bottom quarks by comparing the LO quark-antiquark channel with and without bottom quarks.
Their inclusion allows for diagrams with internal top quarks, giving rise to new topologies, even when bottom quarks are treated as massless.
At LO, these contributions can be studied by including them into \textsc{Powheg-Box-V2} via \textsc{GoSam-3.0}. 

\begin{figure}
\centering
\includegraphics[scale=0.59]{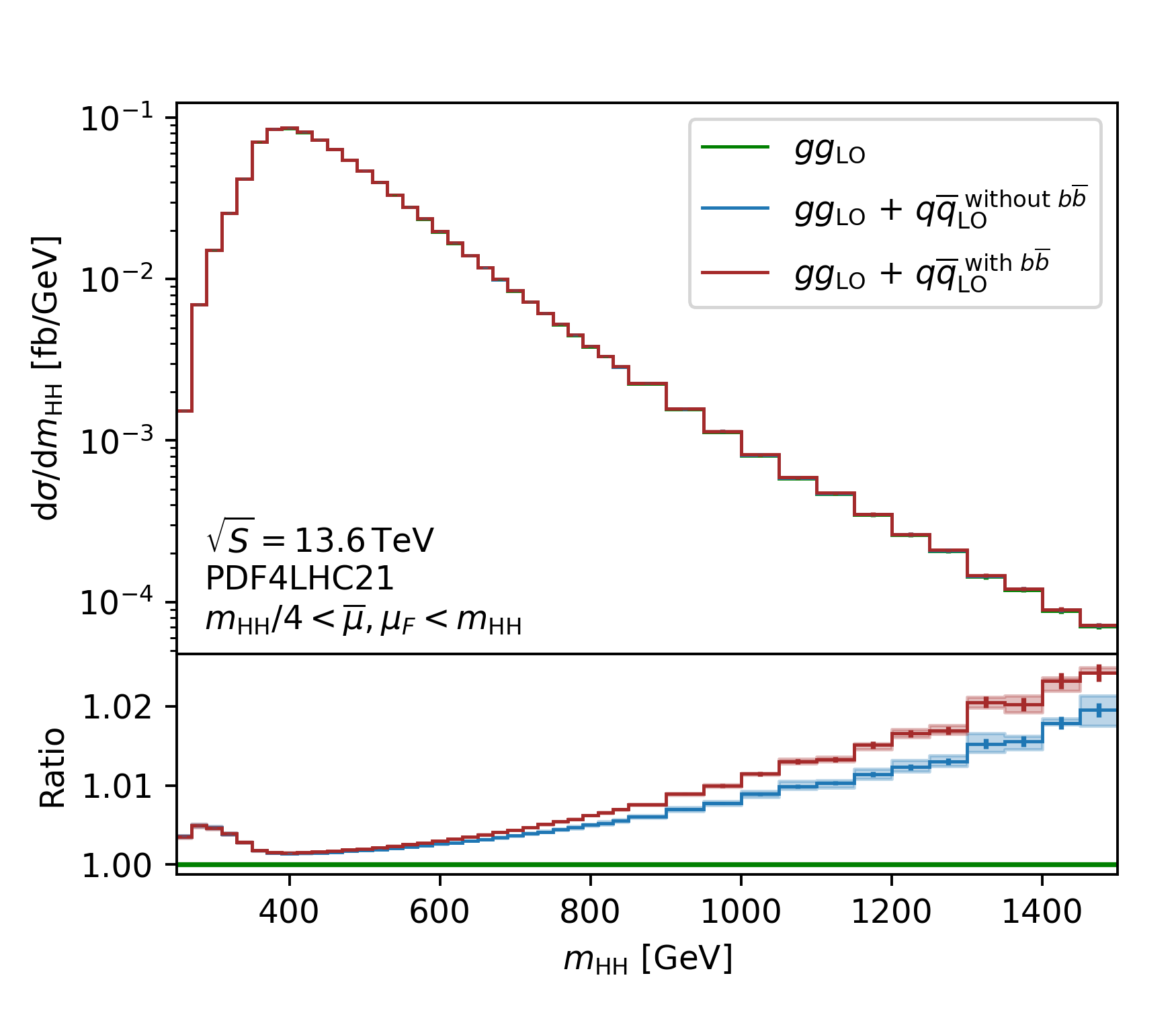}
\includegraphics[scale=0.59]{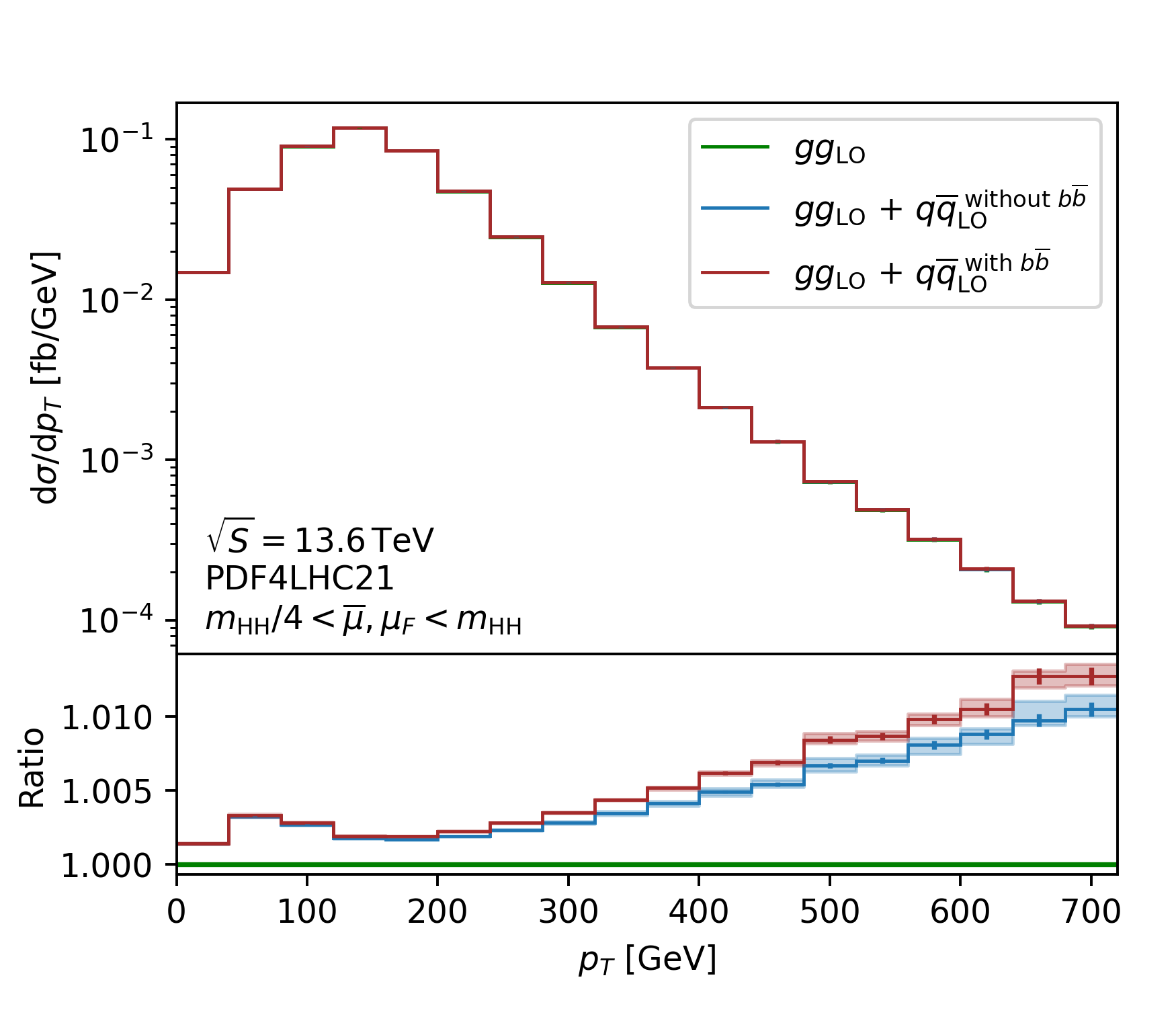}
\caption{Effect of initial state bottom quarks on the invariant mass distribution of the Higgs boson pair (left) and transverse momentum distribution of a single Higgs boson (right). The error bands represent the scale uncertainties of the quark-antiquark channel while the error bars depict the statistical error.}
\label{fig:bottom}
\end{figure}

Figure~\ref{fig:bottom} illustrates the impact of bottom quarks on the invariant mass distribution of the Higgs boson pair (left) and the transverse momentum distribution of a single Higgs boson (right). Including bottom quarks in the initial state leads to a maximum increase of $35\%$ in the high-energy bins, compared to the LO quark-antiquark prediction without bottom quarks. 
At total cross section level, including (massless) bottom quarks in the initial state increases the quark-antiquark channel by about 10\%, because the large-$\mhh$ and large-$p_T$ bins do not contribute much to the total cross section. 
We leave the investigation of the effects of massive bottom quarks in the gluon and quark-antiquark channels to future work, building on results of Ref.~\cite{thesis_DavidBuerg}.

\section{Conclusions}
\label{sec:conclu}

We have calculated the NLO QCD corrections to Higgs boson pair production in the quark-antiquark channel, where the Higgs boson pair at LO is produced through a loop containing electroweak bosons.
Neglecting initial state $b$-quarks as their PDFs are suppressed, the NLO corrections involve two-loop amplitudes with four kinematical variables, including two mass scales, $m_V$ and $m_H$, for which an analytic result has been achieved, using methods similar to the ones presented in Ref.~\cite{Bonetti:2025vfd}.
The one-loop real radiation matrix elements have been generated with \textsc{GoSam-3.0}~\cite{Braun:2025afl}.

The results have been implemented into a Monte Carlo code based on the \textsc{Powheg-Box-V2} framework~\cite{Nason:2004rx,Frixione:2007vw,Alioli:2010xd} to perform phenomenological studies. It turned out that the NLO corrections to the $q\bar{q}$ channel have a K-factor of about $1.6$.
The large K-factor can be understood from contributions at low  $\mhh$ where a hard jet recoils against the Higgs boson pair, and also stems from the fact that the real radiation corrections open up new partonic channels: the $qg$ channel, only present in the real corrections, profits from the large gluon luminosity at the LHC.

While these contributions still amount to less than 1\% of the LO cross section in the gluon channel, the corrections are important at the differential level: they enhance the $\mhh$ distribution close to the Higgs boson pair production threshold by almost 10\% (with $20\,\textup{GeV}$ bins), and also show a Sudakov-type enhancement at high energies.
As the low-$\mhh$ region is very sensitive to modifications of the trilinear Higgs coupling, it is very important to control this region well, and therefore these contributions are not negligible in view of the high-quality data anticipated for the High-Luminosity LHC.

Having addressed the light-quark contributions, the next steps are a detailed study of massive bottom quark effects for Higgs boson pair production, and the combination of the results presented here with corrections of electroweak and QCD type in the gluon channel.




\section*{Acknowledgments}

We would like to thank Jens Braun, Benjamin Campillo, Marius Höfer, Stephen Jones, Matthias Kerner, Pau Petit Ros\`as, Chiara Signorile-Signorile and Augustin Vestner for useful discussions.
M.B.\ wishes to thank Barbara Jäger for her support during the final stages of this project.
This work is supported by the \textit{Deutsche Forschungsgemeinschaft} (DFG, German Research Foundation) under grant no.\ 396021762 - TRR 257, and by the Leverhulme Trust, LIP-2021-014.

\appendix

 \section{Canonical DEQ for \texorpdfstring{\boldmath $\mathcal{A}^{(1)}$}{A^(1)}}
\label{app:A}

In this appendix, we provide the canonical differential equations that obey the 
functions that contribute to $\mathcal{A}^{(1)}$,  
\begin{align}
    \td\vec{\omega}_{1;0} = \td\Omega_{1;0}\,\vec{\omega}_{1;0}\,,
\end{align}
with,
\begin{align}
\vec{w}_{3:0}=
\big\{ w_{1_{0}}^{(0)},w_{1_{1}}^{(1)},w_{2_{1}}^{(1)},w_{5_{1}}^{(1)},w_{7_{1}}^{(1)},w_{2_{2}}^{(2)},
w_{7_{2}}^{(2)},w_{10_{2}}^{(2)},w_{13_{2}}^{(2)},w_{20_{2}}^{\text{(2)}},w_{23_{2}}^{\text{(2)}}\big\}\,,
\end{align}
the only non-vanishing boundary value at $s=0$, 
\begin{align}
w_{1_{0}}^{\text{(0)}}\big|_{s=0} = -1\,,
\end{align}
and with the matrix of coefficients,
\begin{align}
\Omega_{1;0} = &
\left(
\begin{array}{cccccccc}
0 & 0 & 0 & 0 & 0 & 0 & \hdots & 0\\
-2L_{41} & 0 & 0 & 0 & 0 & 0 & \hdots & 0\\
-4L_{10} & 0 & 0 & 0 & 0 & 0 & \hdots & 0\\
2L_{38} & 0 & 0 & 0 & 0 & 0 & \hdots & 0\\
-4L_{11} & 0 & 0 & 0 & 0 & 0 & \hdots & 0\\
0 & -\frac{L_{39}}{4}-\frac{L_{41}}{8} & \frac{L_{21}}{4}-\frac{L_{3}}{4} & 0 & 0 & 0 & \hdots & 0\\
0 & \frac{L_{54}}{4} & 0 & \frac{L_{52}}{4} & 0 & 0 & \hdots & 0\\
0 & -\frac{L_{62}}{2} & -\frac{L_{37}}{2} & \frac{L_{61}}{4} & 0 & 0 & \hdots & 0\\
0 & 0 & 0 & \frac{L_{38}}{4} & 0 & 0 & \hdots & 0\\
0 & -\frac{L_{40}}{4}-\frac{L_{41}}{8} & 0 & 0 & \frac{L_{22}}{4}-\frac{L_{4}}{4} & 0 & \hdots & 0\\
0 & -\frac{L_{70}}{2} & 0 & -\frac{L_{66}}{4} & -\frac{L_{49}}{2} & 0 & \hdots & 0
\end{array}
\right)
\,,
\end{align}
where the integration kernels are, 
\begin{align}
 & L_{3}=\log(t)\,, &  & L_{40}=\log\left(\frac{m_{H}^{2}-r_{5}-2u}{m_{H}^{2}+r_{5}-2u}\right),\\
 & L_{4}=\log(u)\,, &  & L_{41}=\log\left(\frac{m_{H}^{2}-r_{5}-2}{m_{H}^{2}+r_{5}-2}\right),\nonumber \\
 & L_{10}=\log(1-t)\,, &  & L_{49}=\log\left(\frac{s(1-u)-r_{8}}{r_{8}+s(1-u)}\right),\nonumber \\
 & L_{11}=\log(1-u)\,, &  & L_{52}=\log\left(\frac{s\left(2m_{H}^{2}-s\right)-r_{1}r_{4}}{s\left(2m_{H}^{2}-s\right)+r_{1}r_{4}}\right),\nonumber \\
 & L_{21}=\log\left((1-t)m_{H}^{2}+t^{2}\right)\,, &  & L_{54}=\log\left(\frac{(s+8)m_{H}^{2}-2m_{H}^{4}-r_{1}r_{5}-2s}{(s+8)m_{H}^{2}-2m_{H}^{4}+r_{1}r_{5}-2s}\right),\nonumber \\
 & L_{22}=\log\left((1-u)m_{H}^{2}+u^{2}\right), &  & L_{61}=\log\left(\frac{-r_{3}r_{4}-s(st-3t+u)}{r_{3}r_{4}-s(st-3t+u)}\right),\nonumber \\
 & L_{37}=\log\left(\frac{s(1-t)-r_{3}}{r_{3}+s(1-t)}\right), &  & L_{62}=\log\left(\frac{s\left((1-t)m_{H}^{2}+2t\right)-r_{3}r_{5}}{s\left((1-t)m_{H}^{2}+2t\right)+r_{3}r_{5}}\right),\nonumber \\
 & L_{38}=\log\left(\frac{-r_{4}-s+2}{r_{4}-s+2}\right), &  & L_{66}=\log\left(\frac{s(su+t-3u)-r_{4}r_{8}}{r_{4}r_{8}+s(su+t-3u)}\right),\nonumber \\
 & L_{39}=\log\left(\frac{m_{H}^{2}-r_{5}-2t}{m_{H}^{2}+r_{5}-2t}\right), &  & L_{70}=\log\left(\frac{s\left((1-u)m_{H}^{2}+2u\right)-r_{5}r_{8}}{s\left((1-u)m_{H}^{2}+2u\right)+r_{5}r_{8}}\right)\,,\nonumber 
\end{align}
and square roots, 
\begin{align}
r_{1} & =\sqrt{s\left(s-4m_{H}^{2}\right)}\,,\\
r_{3} & =\sqrt{s\left(m_{V}^{4}(s-4m_{H}^{2})+st^{2}-2m_{V}^{2}t(t-u)\right)}\,,\nonumber \\
r_{4} & =\sqrt{s\left(s-4m_{V}^{2}\right)}\,,\nonumber \\
r_{5} & =\sqrt{m_{H}^{2}\left(m_{H}^{2}-4m_{V}^{2}\right)}\,,\nonumber \\
r_{8} & =\sqrt{s\left(m_{V}^{4}(s-4m_{H}^{2})+su^{2}+2m_{V}^{2}u(t-u)\right)}\,.\nonumber 
\end{align}

\bibliographystyle{jhep}
\bibliography{qqHH}

\end{document}